\documentstyle [12pt]{article}
\begin{document}

\begin{titlepage}
hep-ph/9501412
\centerline{SPECTRUM OF ELEMENTARY PARTICLES IN A MODEL OF }
\centerline{HADRON SUPERSYMMETRY.}
\vspace{10mm}

\centerline{S.Kiyanov-Charsky}
\vspace{5mm}
\centerline{Department of Physics,Forestry Academy,}
\centerline{Institutsky pr.5,St-Petersburg 194018,}
\centerline{Russia}
\vspace{15mm}

 {\bf Abstract}
\vspace{5mm}

   We investigate a spectrum of the low-energy composite particles with
 the quantum numbers $J^p=0^\pm,\frac {1}{2}^\pm$ in a $SU_{F}(3)$ model
 of hadron supersymmetry. We derive the mass spectrum of two, three and
 four-quark states and determine all free parameters of a theory,
 including the masses of quarks and diquarks.

\vspace{1in}

PACS: 11.30.Pb,12.40,12.50

\vspace{30mm}
\end{titlepage}

    In the recent years, the supersymmetry (SUSY) approach was applied to
 the low energy physics of hadrons \cite{A}. It was shown that some
 hadronic states form a representation space of the $N=1$ SUSY algebra.
 But unsolved problem is how this algebra acts on the states of
 conventional QCD. In some papers \cite{B}, a possible way to solve this
 problem inside $SU_{F}(3)$ flavour quark model was given, if we identify
 diquarks (colour triplet states of two quarks)as the scalar quark states,
 which are the superpartners of ordinary quarks.

    In this paper, we give a formulation of a model of hadron SUSY and
 derive an appropriate low energy model of hadrons, where SUSY is
 spontaneously broken. We compute the masses of hadrons, using
 experimental data of pseudoscalar two-quark mesons,scalar four-quark
 mesons, and barions \cite{C}. By these data, we determine all free
 parameters of the theory and predict some new states in a scalar meson
 sector.

    Consider a model of hadron SUSY. It contains the following superfields
 (SF's):
 $$\Phi_{\pm\alpha i}=a_{\pm\alpha i}+\sqrt2\theta q_{\pm\alpha i}+
 \theta\theta\phi_{\pm\alpha i} ,$$
 $${\bar \Phi}_{\pm\alpha i}=a^*_{\pm\alpha i}+
 \sqrt2\bar{\theta}{\bar q}_{\pm\alpha i}+
 \bar{\theta}\bar{\theta}\phi^*_{\pm\alpha i}.$$
 Here $$q_{\alpha i}=\pmatrix {q_{+\alpha i}        \cr
                               {\bar q}_{-\alpha i}}  $$
 is a quark dirac spinor field; $a_{\pm\alpha i},a^*_{\pm\alpha i}$ are
 scalar quark fields which can be identified with the diquark fields;
 $\phi_{\pm\alpha i}$, $\phi^*_{\pm\alpha i}$ are auxiliary fields;
 $\alpha$, $i$ are indices of $SU_{c}(3)$ and $SU_{F}(3)$ respectively.
 All these fields are the components of the chiral SF's
 $$\Phi_{\alpha i}=\pmatrix{\Phi_{+\alpha i}       \cr
                            {\bar\Phi}_{-\alpha i}}  . $$
 We use notations of ref.\cite{D}. Let us include some additional external
 Higgs SF's $\chi,\bar{\chi}$ and gauge SF's $u$, $v$ in an adjoint
 representation of the chiral groups $SU_{RF}(3)$ and $SU_{LF}(3)$,
 respectively. These SF's are needed for soft SUSY breaking, they also give
 masses for quarks and diquarks.  In the component form we have
  $$u=-v=\frac {1}{4}\theta\theta\bar{\theta}\bar{\theta}D;$$
  $$\chi=\frac{1}{2}(m+\theta\theta f) ,\bar{\chi}=\frac{1}{2}
  (m+\bar{\theta}\bar{\theta}f) ,$$
 and $D,f,m$-are $SU_{F}(3)$ matrices of the form
  $$\pmatrix { m_u& 0 & 0   \cr
                0 &m_d& 0   \cr
                0 & 0 &m_s  }  . $$

 The action of the theory is $S_0=<\Phi,{\bf K_0}\Phi>$ (all indices
 are omitted), where we use a pairing in a chiral SF space
 $<\Phi_1,\Phi_2>=\int d^{4}xd^{2}\theta\Phi_{+1}\Phi_{-2}+ \int
 d^{4}xd^{2}\bar{\theta}{\bar\Phi}_{+2}{\bar\Phi}_{-1}$, instead of usual
 pairing for the dirac spinor fields $<\psi_{1},\psi_{2}>=
 \int d^{4}x\psi_{+1}\psi_{-2}+\int d^{4}x{\bar\psi}_{+2}{\bar\psi}_{-1}$.
 The superkinetic operator ${\bf K_0}$ is
 $${\bf K_0}=\pmatrix {   \chi       & \bar{D}^{2}e^{-v} \cr
                        D^{2}e^{u}  &       \bar{\chi}   }    $$
 in a superfield form, and

 $${\bf K_0}=\pmatrix { \frac{m}{2}     &  1  &  0  &    0           &
      0                  &     0            \cr
                       \partial^2+\frac{D}{4}  &  0  &  0  &\frac{f}{2} &
      0                  &     0            \cr
                    \frac{f}{2}     &  0  &  0  &\partial^2+\frac{D}{4} &
      0                  &     0            \cr
                             0          &  0  &  1  &\frac{m}{2}     &
      0                  &     0            \cr
                             0          &  0  &  0  &    0           &
  \frac{m}{2}             & i\hat{\partial} \cr
                             0          &  0  &  0  &    0           &
   i\hat{\bar{\partial}} &\frac{m}{2}       } $$
 in a component form; it is acting on a component field graded space
 $$\Phi=\pmatrix{   a_{+}    \cr
                 \phi_{+}    \cr
                    a^*_{-}  \cr
                 \phi^*_{-}  \cr
                    q_{+}    \cr
                 {\bar q}_{-} }  . $$

   Now we introduce an effective low energy action in a usual way \cite{E}.
 We consider a chiral transformation matrix $\widetilde{\Omega}$ which is
 similar to ordinary matrix $\Omega=e^{i\gamma_{5}\pi(x)}$ for the chiral
 field $ \pi(x)$. Instead of $\pi(x)$ we have the SUSY extension:
 $$\gamma=\pi+i\sigma+\sqrt{2}\theta b+\theta\theta F, $$
 $$\bar{\gamma}=\pi-i\sigma+\sqrt{2}\bar{\theta}\bar{b}+
 \bar{\theta}\bar{\theta}F^* , $$
 where $\gamma,\bar{\gamma}$ are in an adjoint representation of
 $SU_{F}(3)$, and $\pi$, $\sigma$, $b$, $\bar{b}$ are stand for
 pseudoscalar, scalar and spinor effective fields, respectively.
 So, for the supermatrix $\widetilde{\Omega}$ we have:
 $$\widetilde{\Omega}=\pmatrix {e^{i\gamma_{2}} &    0          &
  0         &      0         &       0      &        0          \cr
                                   0        &e^{i\gamma_{2}}&
  0         &   -iF^*        &       0      &   i\bar{b}        \cr
                                  iF        &    0          &
e^{-i\gamma^*_2}&      0     &      -ib     &        0          \cr
                                   0        &    0          &
  0         &e^{-i\gamma^*_2}&       0      &        0          \cr
                                  ib        &    0          &
  0         &      0         &e^{i\gamma_1} &        0          \cr
                                   0        &    0          &
  0         &    -i\bar{b}   &       0      &e^{-i\gamma^*_1}   }  .  $$
 Here $\gamma_1=\pi_1+i\sigma_1,\gamma^*_1=\pi_1-i\sigma_1$ are parameters
 of chiral transformation of the pure quark fields, so $\pi_1$ and
 $\sigma_1$ - are pseudoscalar and scalar two-quark effective states,
 as in ordinary QCD (formally $\gamma_1=q\bar{q}$). Parameters
 $\gamma_2=\pi_2+i\sigma_2,\gamma^*_2=\pi_2-i\sigma_2$, are acting on the
 pure bosonic diquark states, so $\pi_2$ and $\sigma_2$ can be
 identified as pseudoscalar and scalar four-quark states (formally
 $\gamma_2=aa^*=qq\bar{q}\bar{q}$). And the fermionic parameters $b$,
 $\bar{b}$ stand for barion states ($b=a^*q=qqq$).

    Then, as usual,
 $$\Gamma_{eff}(\pi_1,\sigma_1,\pi_2,\sigma_2,b,\bar{b})=
 \frac{1}{2}S{\rm Tr}\log{\bf K}=
 -\frac{1}{4}\int^{\frac{1}{\mu^2}}_{\frac{1}{\Lambda^2}}
 \frac{dt}{t}S{\rm Tr}e^{-t{\bf K}^2} ,$$
 where ${\bf K}=\widetilde{\Omega}{\bf K_{0}}\widetilde{\Omega}$, and we
 have used the proper time formalizm for definition of
 $S{\rm Tr}\log{\bf K}$. After the computation of supertrace and all
 needed integrals in ${\rm Tr}$,we obtain the desired quadratic
 part of $\Gamma_{eff}$:
\begin{eqnarray}
 \Gamma_{eff} & = & 4{\rm tr} q^2m^2\int d^4x \pi_2 \partial^2 \pi_2+
 4{\rm tr}(p^2+q^2m^2)
 \int d^4x(b{\bar{\sigma}}^{\mu}i\partial_{\mu}\bar{b})+\nonumber\\
 & &2{\rm tr}(2p^2+q^2(m^2-D))
 \int d^4x\sigma_2 \partial^2 \sigma_2-\nonumber\\
 & &{\rm tr}(4f^2q^2(1+\frac{4m^2q^2}{p^2+q^2m^2})-
 4D(p^2+2q^2m^2)-m^2(6p^2+q^2m^2))\times\nonumber\\
 & &\int d^4x\sigma^2_2-2{\rm tr}q^2mf
 \int d^4x(bb+\bar{b}\bar{b})-\nonumber\\
 & &2{\rm tr}(2q^2m^2(-D)-m^2(2p^2+q^2m^2))
 \int d^4x\pi^2_2+4{\rm tr}p^2\int d^4x\pi_1 \partial^2 \pi_1+\nonumber\\
 & &2{\rm tr}q^2m^2\int d^4x\sigma_1 \partial^2 \sigma_1-
 2{\rm tr}m^2(2p^2+q^2m^2)\int d^4x\pi^2_1-\nonumber\\
 & &{\rm tr}m^2(6p^2+q^2m^2)\int d^4x\sigma^2_1 .\nonumber
\end{eqnarray}

 Here $p^2,q^2$-are functions of $\Lambda^2$, $\mu^2$, and they determine
 the low-energy region of integration.

   $\Gamma_{eff}$ is independent of regularization procedure because it is
 expressed through the Seeley coefficients for the operator {\bf K}(for
 the review, see ref.\cite{F}).

   After determination of all parameters $p^2$, $q^2$, $m$, $f$, $D$ by the
 data of meson and barion spectroscopy, we have the following table for the
   composite low-energy  particles.\\

\vspace{5mm}
\begin{tabular}{|c|c|c|c|c|c|c|} \hline
\multicolumn{7}{|c|}{{\it Two-quark meson states}}\\ \hline
\multicolumn{3}{|c|}{$ J^p=0^-$ }& &
\multicolumn{3}{|c|}{$ J^p=0^+$ }\\ \hline
     mesons & experiment & theory &        &mesons & experiment &
 theory \\
            &  (MeV)     &  (MeV) &        &       & (MeV)    &
  (MeV) \\ \hline
$\pi^{\pm,0}$&    140    &  140   &  $I=1$ &         -  &   - &
    370 \\ \hline
$ K^{\pm,0}$ &    490    &  490   &$I=\frac{1}{2}$&  -  &  -  &
    860 \\ \hline
$\eta_0$     &    550    &  560   & $I=0$  & $f_0$      &  975&
    970 \\ \hline
\end{tabular}
\\
\vspace{10mm}
\\

\begin{tabular}{|c|c|c|c|c|c|c|} \hline
\multicolumn{7}{|c|}{{\it Four-quark meson states}}\\ \hline
\multicolumn{3}{|c|}{$ J^p=0^-$ }& &
\multicolumn{3}{|c|}{$ J^p=0^+$ }\\ \hline
     mesons & experiment & theory &        &mesons & experiment &
 theory \\
            &  (MeV)     &  (MeV) &        &       & (MeV)    &
  (MeV) \\ \hline
  $\pi$     &   1300     & 1350   & $I=1$  & $ a_0$&  980     &
  980   \\ \hline
     -      &     -      & 1500   &$I=\frac{1}{2}$& $K^*_0$ & 1430 &
   1430 \\ \hline
  $\eta$    &   1440     & 1550   & $I=0$  & $ f_0$ & 1590    &
 1550   \\ \hline
\end{tabular}
\vspace{5mm}

   For the sake of simplicity, all theoretical masses of the barion octet
 $J^p=\frac{1}{2}^{\pm}$ are equal to the mass of $\Sigma^{\pm}$:
 $m_{b,\bar{b}} = m_{\Sigma}=1190 MeV$.

   In our model, we use the octet of scalar mesons as pure four-quark
 states for definition of the matrix $D$, barion octet for determination
 of the matrix {\it f}, and the octet of pseudoscalar mesons as pure
 two-quark states for determination of the quark mass matrix $m$.
 Then, we predict the masses of pure two-quark scalar states and pure
 four-quark pseudoscalar states, as shown at the tables.

   Finally, we get the quark masses
 $$m_u=m_d=95 MeV, m_s=240 MeV,$$
 and the diquark masses
 $$m^d_u=m^d_d=(320,1020) MeV, m^d_s=(540,1070) Mev.$$

   In this paper, we do not discuss a mixing between two, four-quark and
 glueball states in the sector of scalar mesons. It is related to the
 theories of non- $q\bar{q}$ states(for the review, see \cite{G} and
 references therein). The next step of the investigation would be to take
 into account a mixing between such states in a real world of scalar
 mesons.

\vspace{10mm}
 {\sl Acknowledgements.} The author would like to thank Dr. Vassilevich D.
 for very helpful discussions.

\end{document}